\documentstyle[11pt,newpasp,twoside,epsf]{article}
\begin{document}
\title{Low-Luminosity AGNs and Unification}

\author{Aaron J. Barth}
\affil{Harvard-Smithsonian Center for Astrophysics, 60 Garden St.,
 Cambridge, MA 01238}

\begin{abstract}
More than one third of all nearby galaxies show indications of
low-luminosity nuclear activity.  These low-luminosity AGNs (LLAGNs)
are traditionally classified into a variety of categories depending on
their forbidden-line ratios and on the presence or absence of
broad-line emission.  By analogy with Seyfert unification models, it
is natural to ask whether the various LLAGN types are different
manifestations of the same underlying phenomenon, with observed
differences being solely due to orientation and obscuration, or
whether there are fundamentally distinct physical processes at work in
different categories of LLAGNs.  This contribution reviews some recent
observations of LLAGNs in the context of AGN unification scenarios.
\end{abstract}

\section{Introduction}

During the past two decades, unification models have been remarkably
successful at clarifying the underlying connections between a variety
of types of AGNs, including Type 1 and 2 Seyferts, radio galaxies and
quasars (Antonucci 1993).  These unification scenarios attempt to
explain the differences between certain AGN classes as the result of
varying orientation to our line of sight of the central engine and an
``obscuring torus'' that surrounds it.  This review explores the
question of whether similar unification ideas can also be applied to
low-luminosity AGNs (LLAGNs), which make up the vast majority of the
AGN population.  These include low-luminosity Seyferts, low-ionization
nuclear emission-line regions (LINERs), and transition-type objects
with properties intermediate between those of LINERs and those of
normal H~II nuclei.

Unification studies of Seyfert galaxies are primarily concerned with
figuring out whether Seyfert 2s are intrinsically the same kind of
AGNs as Seyfert 1s.  For LINERs, the unification question takes on a
somewhat different focus, because it is not clear whether all of the
Type 2 LINERs and transition objects are genuine AGNs at all.
Ultimately, we may be able to achieve a partial unification in which
some Type 2 objects are shown to be obscured or very faint
accretion-powered sources, while others may prove to be completely
unrelated to the AGN phenomenon.  This has important ramifications for
AGN demographics, because the population of LINER 2s and transition
objects outnumbers all other types of AGNs combined.  The fraction of
galaxies containing genuine AGNs also sets a lower limit to the
fraction of galaxies containing supermassive black holes.  This is
still a pertinent issue because dynamical measurements of black hole
masses have been performed only for a few dozen galaxies, while AGN
surveys can indirectly detect black holes in vastly more galaxies, and
over a broader range of Hubble types.

In applying unification ideas to LLAGNs, several questions arise.  To
what extent is our view of LLAGNs determined by orientation and
obscuration?  Do Type 2 LLAGNs show evidence for obscuring tori in the
form of polarized broad-line emission, heavily absorbed X-ray sources,
or ionization cones?  And are some objects classified as LLAGNs
actually powered predominantly or entirely by stellar processes,
rather than by accretion?
  
Other aspects of the LLAGN phenomenon will be covered in more detail
in other contributions to this volume, including those by Nagar (radio
observations), Mushotzky (X-ray observations), and Ho (spectral energy
distributions and central engine physics).  Additional recent reviews
of the properties of LLAGNs, including discussion of unification
issues, are given by V\'eron-Cetty \& V\'eron (2000) and by Ho (2002).

\section{LLAGN Classification and Demographics}

\subsection{Optical Classification of LLAGNs}

The most complete and comprehensive surveys for nearby AGNs have been
carried out in the optical, and emission-line nuclei are classified
into a few general categories based on their forbidden-line ratios.
The basic categories include H~II nuclei (i.e., nuclei whose emission
lines are predominantly powered by young, massive stars), Seyferts,
and LINERs.  Heckman (1980) first defined LINERs as galaxies whose
spectra satisfy [O~II] $\lambda3727$ / [O~III] $\lambda5007 \geq 1$
and [O~I] $\lambda6300$ / [O~III] $\lambda5007 \geq 1/3$.  Other
authors have often used equivalent classifications based on
[O~I]/H$\alpha$ and [O~III]/H$\beta$, since these ratios are less
sensitive to reddening.  The exact cutoff between LINERs and Seyferts
is essentially arbitrary, however.  There is also a category of
``transition-type'' nuclei, whose emission-line ratios are
intermediate between those of H~II regions and LINERs.  Transition
nuclei are sometimes referred to as weak-[O~I] LINERs (Filippenko \&
Terlevich 1992), because their spectra differ from LINERs mainly in
that their [O~I] $\lambda6300$ emission is too weak to meet the LINER
classification criteria.

Low-ionization nebulae satisfying the LINER definition occur in a
variety of environments, including the nuclei of predominantly
early-type (E--Sb) galaxies, superwind galaxies, filaments of gas in
cooling flows, and some ULIRGs.  The focus of this review will be on
the first category: nearby galaxies with low-ionization emission in
the inner $r\la200$ pc.  These LINERs typically have bolometric
luminosities of $\la10^{42}$ erg$^{-1}$ s$^{-1}$ (Ho 1999), so they
are orders of magnitude less luminous than powerful Seyferts and QSOs.
Many nearby radio galaxies such as M87 and M84 are LINERs, and in fact
LINERs as a class appear to be radio-loud objects, even those in
spiral host galaxies (Ho 1999).  The survey of Ho et al.\ (1997a)
found that 11\% of nearby galaxies are Seyferts, 19\% are LINERs, and
13\% have LINER/H~II transition nuclei.  The transition nuclei are
often considered to be a subset of the LINERs, although recent data
suggests that they may be unrelated phenomena (see \S5).

The low luminosity of these objects makes them difficult targets for
observational study, even in very nearby galaxies.  For example, it is
essentially impossible to discern whether LINERs have nonstellar,
featureless continua from ground-based observations, since their
nuclear spectra are dominated by starlight from the surrounding galaxy
bulge.  The optical narrow emission lines of LINERs can be observed
without much difficulty, but their interpretation has long been a
source of controversy because the optical line ratios can be
reproduced reasonably well by models based on a variety of different
physical mechanisms, including shock heating (Dopita \& Sutherland
1996), photoionization by a nonstellar continuum (Ferland \& Netzer
1983; Halpern \& Steiner 1983), or photoionization by hot stars
(Shields 1992).  Thus, it is important to look for other signs of
nonstellar activity in these objects.

\subsection{Broad Emission Lines in LLAGNs}

Some recent reviews of AGN emission lines have categorically denied
the very existence of broad emission lines in LINERs as a class
(Krolik 1999; Sulentic, Marziani, \& Dultzin-Hacyan 2000).  However,
there is no doubt that many LINERs do indeed have broad-line regions
(BLRs); this section briefly reviews the evidence.

Following Heckman's pioneering survey, the detailed study of M81 by
Peimbert \& Torres-Peimbert (1981) and the spectroscopic surveys by
Stauffer (1982), Keel (1983), and Filippenko \& Sargent (1985) showed
that in some objects the H$\alpha$ line has a broad component or broad
wings which are not present on the forbidden-line profiles, indicating
that the broad emission originates in a physically distinct region.
Detecting broad-line emission in LLAGNs is tricky, because the broad
lines are faint and the [N II] $\lambda\lambda$6548, 6583 lines are
superposed on the wings of H$\alpha$.  (Broad H$\beta$ would be a
cleaner measurement since it is less contaminated by blending, but it
is usually too faint to detect in ground-based spectra.)  The
underlying starlight continuum must be subtracted carefully prior to
fitting the profiles, to remove the effects of stellar absorption
features.  Also, the stellar continua of galaxy bulges have a bump
centered roughly around the H$\alpha$+[N~II] blend which can be
confused with broad-line emission if it is not subtracted properly.

\begin{figure}[t]
\plottwo{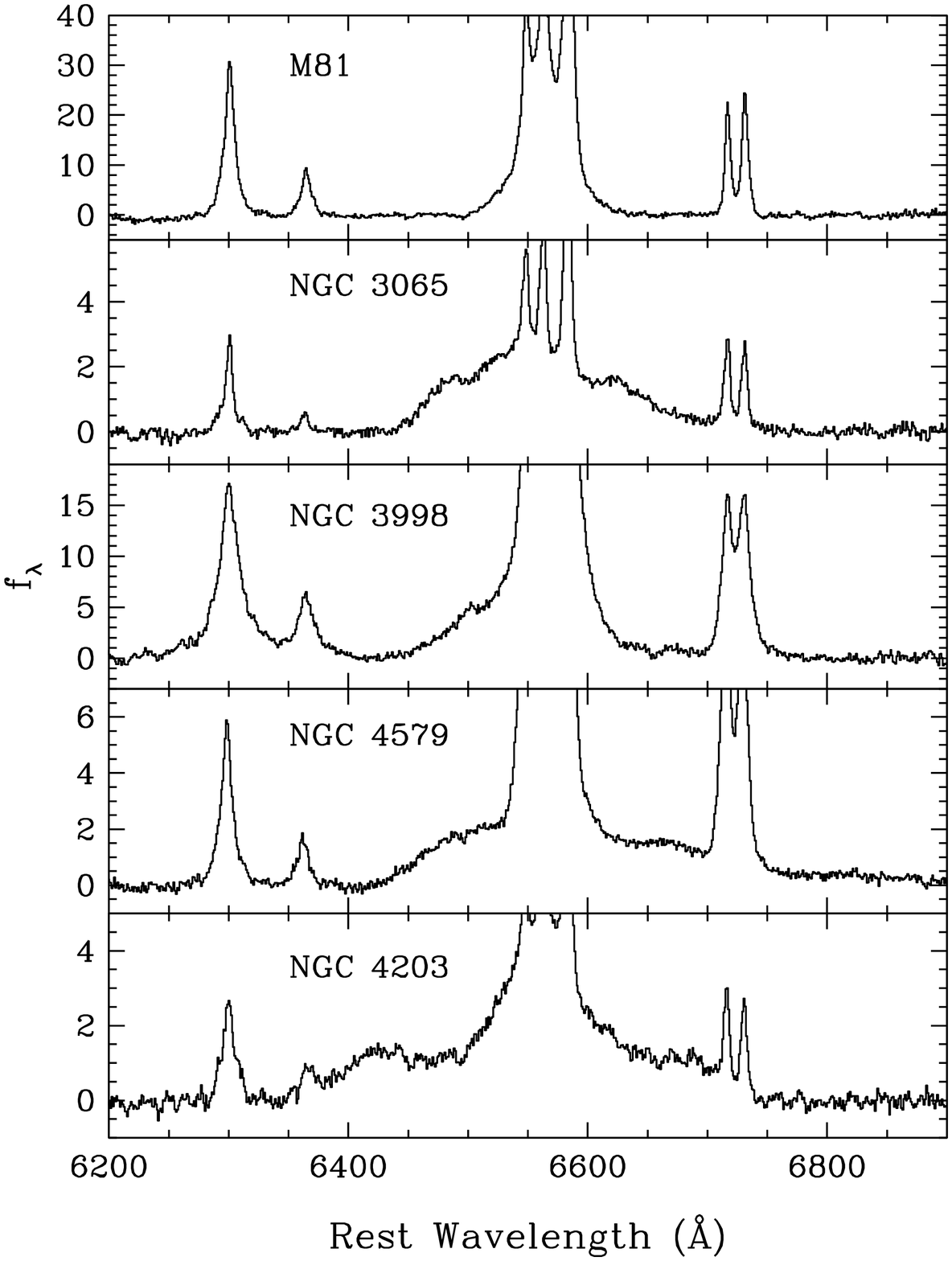}{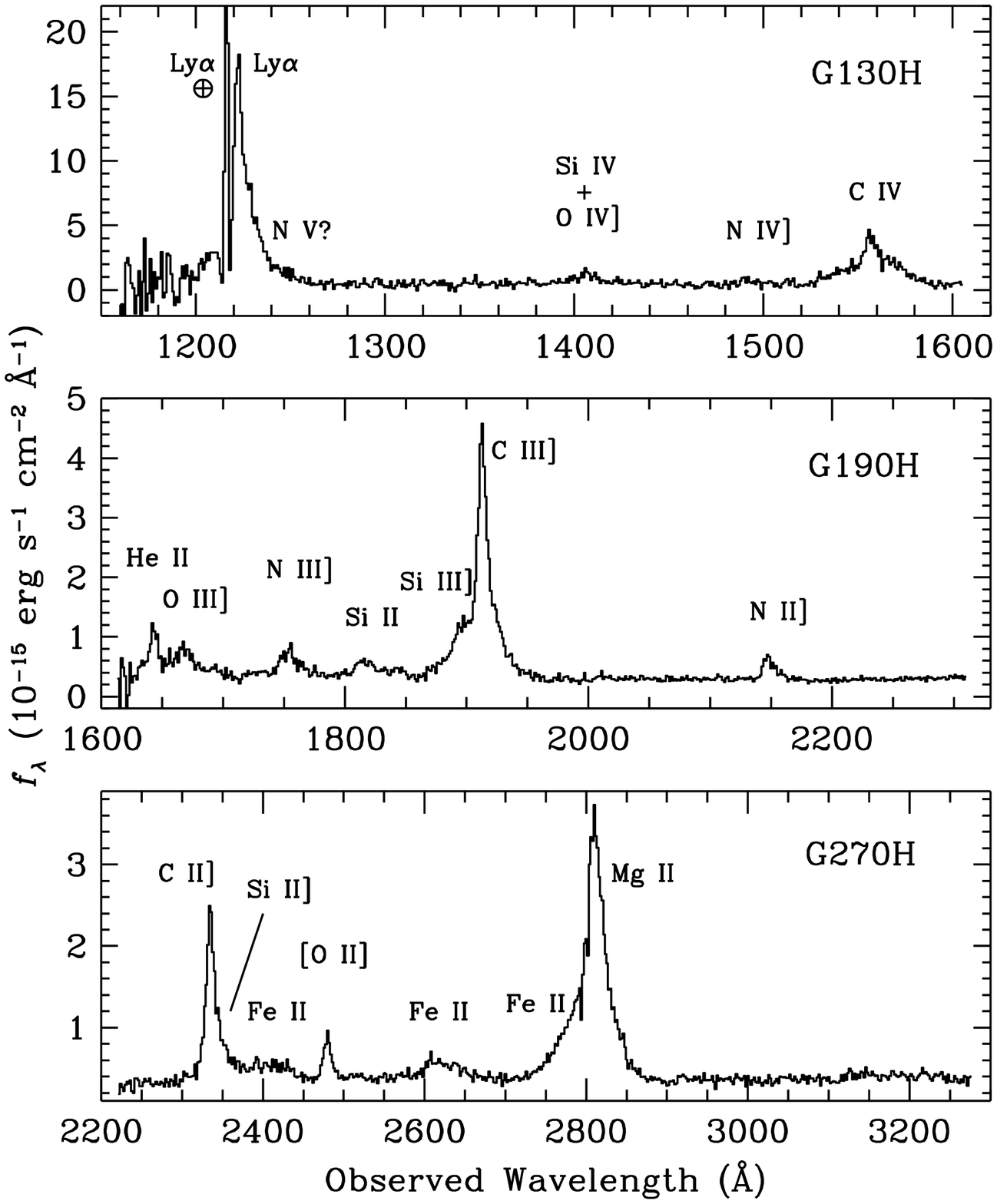}
\caption{Examples of broad emission lines in LINERs.  \emph{Left:}
Starlight-subtracted spectra of LINER 1s, obtained in March 2001 at
the MMT Observatory.  \emph{Right:} \emph{HST} UV spectrum of NGC
4579, from Barth et al.\ (1996), showing broad components of
Ly$\alpha$, C~IV, C~III, and Mg~II.}
\end{figure}

The most comprehensive ground-based survey is that of Ho et al.\
(1997a,b).  They performed decompositions of the H$\alpha$+[N~II]
blend using the [S~II] lines as templates for the forbidden-line
profiles.  Overall, 20\% of the LLAGNs required a broad component to
fit the H$\alpha$ emission, with a median FWHM of 2200 km s$^{-1}$ and
median luminosity of $10^{39}$ erg s$^{-1}$.  The figure of 20\%
should be considered a lower limit, because of the difficulty of
detecting very faint broad lines.  If a similar survey were done with
\emph{HST} and a much narrower slit, it would most likely find a
larger fraction of Type 1 objects.  An intriguing recent development
has been the discovery of extremely broad (FWZI $\approx
15,000-20,000$ km s$^{-1}$) double-peaked or double-shouldered
H$\alpha$ emission lines in several LINERs; see Ho et al.\ (2000) for
a summary of their properties. From the number of double-peaked
emitters detected so far (some of which have been transient features),
the fraction of LINER 1s having double-peaked emission might be of
order $\sim10\%$, but the actual frequency is unknown.  I have
recently begun a ground-based spectroscopic survey of LINER 1s to try
to address this question; the spectra of a few objects with unusual
broad-line profiles are shown in Figure 1.

Ultraviolet (UV) spectra of a few LINER 1s have revealed additional
broad emission lines, confirming the Type 1 classification of these
objects. However, only a handful of LINER 1s have been observed
spectroscopically in the UV, first with \emph{IUE} (Peimbert \&
Torres-Peimbert 1981; Reichert et al.\ 1992) and later with \emph{HST}
(Ho et al.\ 1996; Barth et al.\ 1996).  Figure 1 shows one example,
NGC 4579, in which the broad C~IV line has FWHM = 6600 km s$^{-1}$,
comparable to the linewidths seen in classical, luminous Seyfert 1s.  

The LINER 1s that have been studied in detail almost invariably show
very clear AGN-like properties, such as compact flat-spectrum radio
cores and compact hard X-ray sources (Ho 2002). Their spectral energy
distributions show that the central engines are broad-band emitters
similar in many respects to luminous AGN, albeit with some clear
systematic differences that suggest a different structure for the
central engine (see Ho, this volume).  The accumulated evidence leaves
essentially no doubt that LINER 1s are a category of accretion-powered
AGNs. The status of the LINER 2s is less clear, however, and this is
the focus of the unification question for LLAGNs.

\textbf{A note on classification:} Almost all broad-lined LLAGNs are
classified as type 1.8--1.9 on the Osterbrock (1981) system,
indicating that broad H$\alpha$ is weakly but definitely visible,
while broad H$\beta$ is either extremely weak or not detected.  X-ray
observations of several type 1.8--1.9 LLAGNs have shown that most of
these objects are not heavily obscured, so that we have a clear view
of the central engine.  For example, the LINER 1.9 NGC 4579 has a low
obscuring column of $N_H \approx 4\times10^{20}$ cm$^{-2}$ (Terashima
et al.\ 1998), consistent with the fact that broad lines are clearly
seen in its UV spectrum.  For such objects, the type 1.8--1.9
classification is primarily the result of starlight contamination,
which dilutes the equivalent width of broad H$\alpha$ and H$\beta$;
many of these would probably be classified as type 1.5 if they were
observed through a spectroscopic aperture small enough to exclude the
surrounding starlight.  On the other hand, some type 1.9 LLAGNs are
very heavily obscured sources in which the BLR must be substantially
or completely hidden from direct view.  Well-known examples include
NGC 4258 (Makishima et al.\ 1994) and NGC 1052 (Guainazzi \& Antonelli
1999; Weaver et al.\ 1999); in such objects the faint broad-line
emission may be seen in reflected light.

Thus, the optical classification of LLAGNs into decimal subtypes is by
itself not a good indicator of the degree of obscuration of the
nucleus; it appears to be a function of starlight dilution in some
objects, but obscuration in others.  Some caution is warranted if
LLAGNs of type 1.8--1.9 are included with higher-luminosity type
1.8--1.9 Seyferts in statistical studies of X-ray absorption or other
properties (e.g., Risaliti, Maiolino, \& Salvati 1999).

\section{HST Imaging Surveys}

High-resolution imaging of LINERs with \emph{HST} has provide
important constraints on the nature of the central engines and on the
location and extent of obscuring material.  UV imaging surveys by Maoz
et al.\ (1995) and Barth et al.\ (1998) found nuclear UV emission (at
$\sim2100$ \AA) in $\sim25\%$ of the LINERs that were observed.  About
half appear pointlike at the resolution of \emph{HST} and thus are
good candidates for being genuine LLAGNs with nonstellar continua.
Barth et al.\ (1998) showed that the low UV detection rate is
primarily due to dust obscuration of the nuclei.  The UV-dark LINERs
are systematically found in higher-inclination host galaxies than the
UV-detected LINERs, and the UV-detected galaxies also have lower
reddening as measured from the H$\alpha$/H$\beta$ ratio.  This
suggests that the predominant obscuring structures are foreground dust
lanes that are preferentially aligned with the host galaxy disks.
Similar 100-pc scale obscuring structures may be present in Seyfert
galaxies as well (Maiolino \& Rieke 1995; Malkan, Gorjian, \& Tam
1998).  Consistent with this hypothesis, \emph{HST} $V$-band images
revealed optically thick dust lanes in a substantial fraction of the
sample galaxies, particularly in the UV-dark objects (Figure 2).
Thus, the majority of LINERs probably do have UV sources in their
nuclei (which could be either AGNs or young star clusters), but in
most cases the UV source lies behind enough dust to render it
invisible in a 1-orbit \emph{HST} exposure.

\begin{figure}[t]
\plottwo{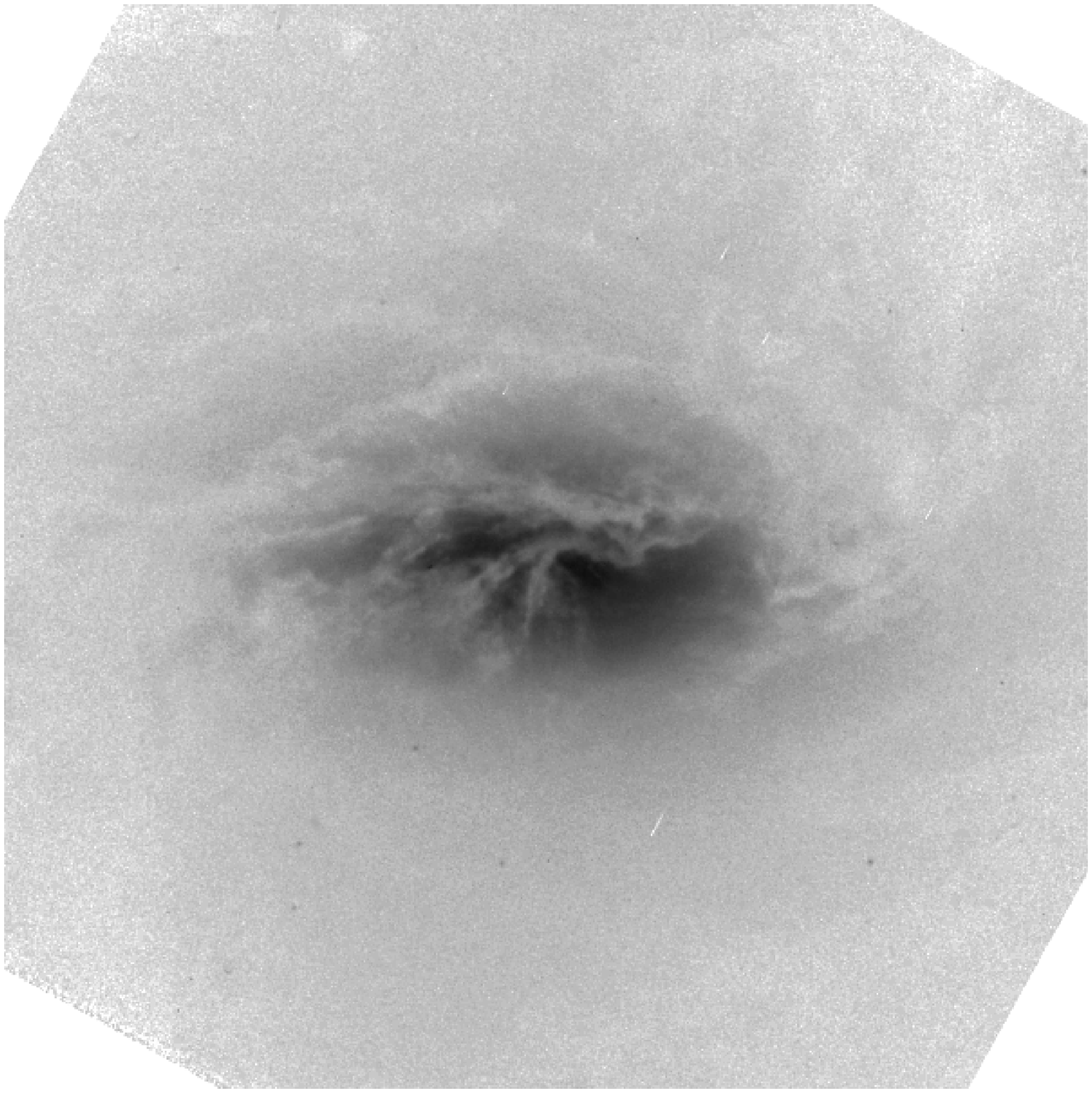}{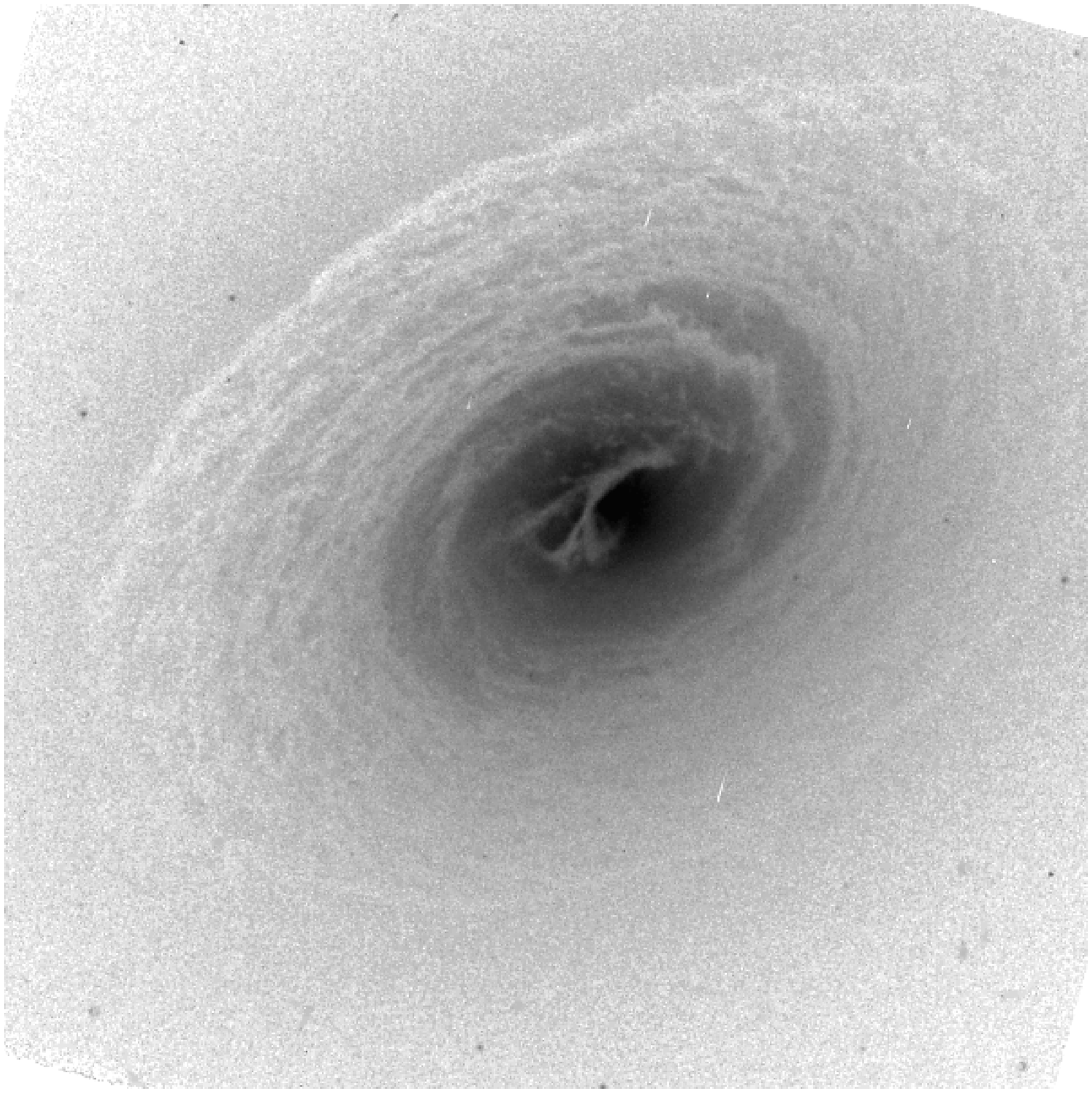}
\caption{\emph{HST} $V$-band images of dusty LINER 2 nuclei, from
Barth et al.\ (1998).  Pictured are the central regions of the S0
galaxies NGC 3166 (left) and NGC 3607 (right).  Image size is
$27\arcsec\times27\arcsec$.
\label{hstimages}}
\end{figure}

Pogge et al.\ (2000) obtained \emph{HST} narrow-band [O~III] and
H$\alpha$+[N~II] images of 14 LINERs to study the narrow-line region
(NLR) morphology and search for ionization cones like those found in
some Seyfert 2 galaxies.  A cone-like NLR morphology was found in only
one object, NGC 1052, and possibly detected in another, M84.  In
general, the NLRs of LINERs appear irregular, with most objects
showing clumpy knots and filaments, and overall there are no clear
differences between type the NLRs of 1.9 and type 2 LINERs.  Pogge et
al.\ also examined broad-band color maps and unsharp-masked images to
search for dust lanes.  Most of the UV-dark LINERs were found to have
dust lanes in the immediate environment of the nucleus, further
supporting the hypothesis that foreground dust plays an important role
in blocking our view of the central engines.

\section{Spectropolarimetry of LLAGNs}
\label{pol}

The discovery of a hidden BLR in polarized light in the spectrum of
NGC 1068 (Antonucci \& Miller 1985) was the key observation that
proved that at least some Seyfert 2s are really Seyfert 1s in which
the nuclear continuum and BLR are obscured.  Following this discovery,
a few groups attempted spectropolarimetric observations of LINERs, to
test whether any LINER 2s contained hidden Type 1 nuclei and whether
the broad H$\alpha$ components in LINER 1s are seen in direct or
scattered light.  This proved to be an impossible task for 3--4 meter
telescopes, as the overwhelming dominance of starlight in the optical
spectra pushes any polarization signature to extremely low levels
($<1\%$ in the continuum).  At these low levels, it is difficult to
discern whether any detected continuum polarization is due to
scattering of nonstellar radiation, or instead merely the result of
foreground dust in the host galaxy imprinting a polarization signature
on the starlight spectrum.

Wilkes et al.\ (1995) were the first to successfully detect continuum
and emission-line polarization from a LLAGN, the Seyfert 1.9 nucleus of
NGC 4258.  NGC 4258 is well known for the H$_2$O maser emission from
its rotating, edge-on, circumnuclear disk that surrounds a
$4\times10^7$ M$_\odot$ black hole (Miyoshi et al.\ 1995).  The narrow
emission lines are very strongly polarized ($1-10\%$), and the
polarization vector of the emission lines and continuum is almost
exactly parallel to the disk plane.  The emission lines are broader in
polarized light than in the total-flux spectrum, so scattering (rather
than transmission through aligned dust grains) is the most viable
polarization mechanism, and provides a consistent explanation for the
observed angle of polarization.  The continuum polarization is only
0.2\% because the nuclear spectrum is dominated by starlight, but
imaging polarimetry showed that region emitting the polarized light is
compact and centered on the nucleus.  NGC 4258 is one of the few
objects in which a clear connection can be made between polarized
nuclear emission and the presence of an edge-on obscuring structure on
subparsec scales.  This provides an important confirmation of the
Seyfert unification concept, and shows that the obscuring torus model
can apply to objects with low luminosity.

\begin{figure}[t]
\plottwo{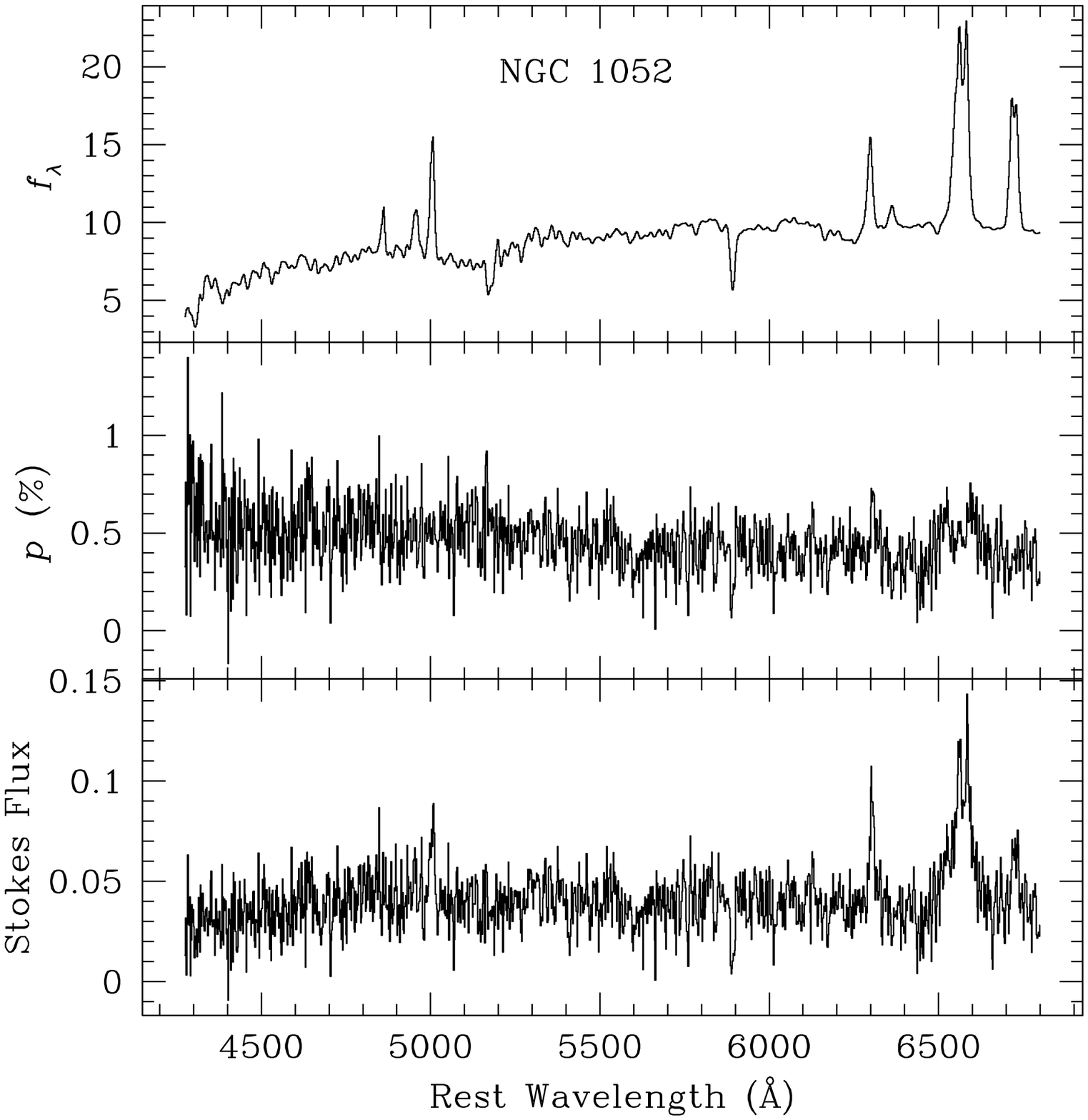}{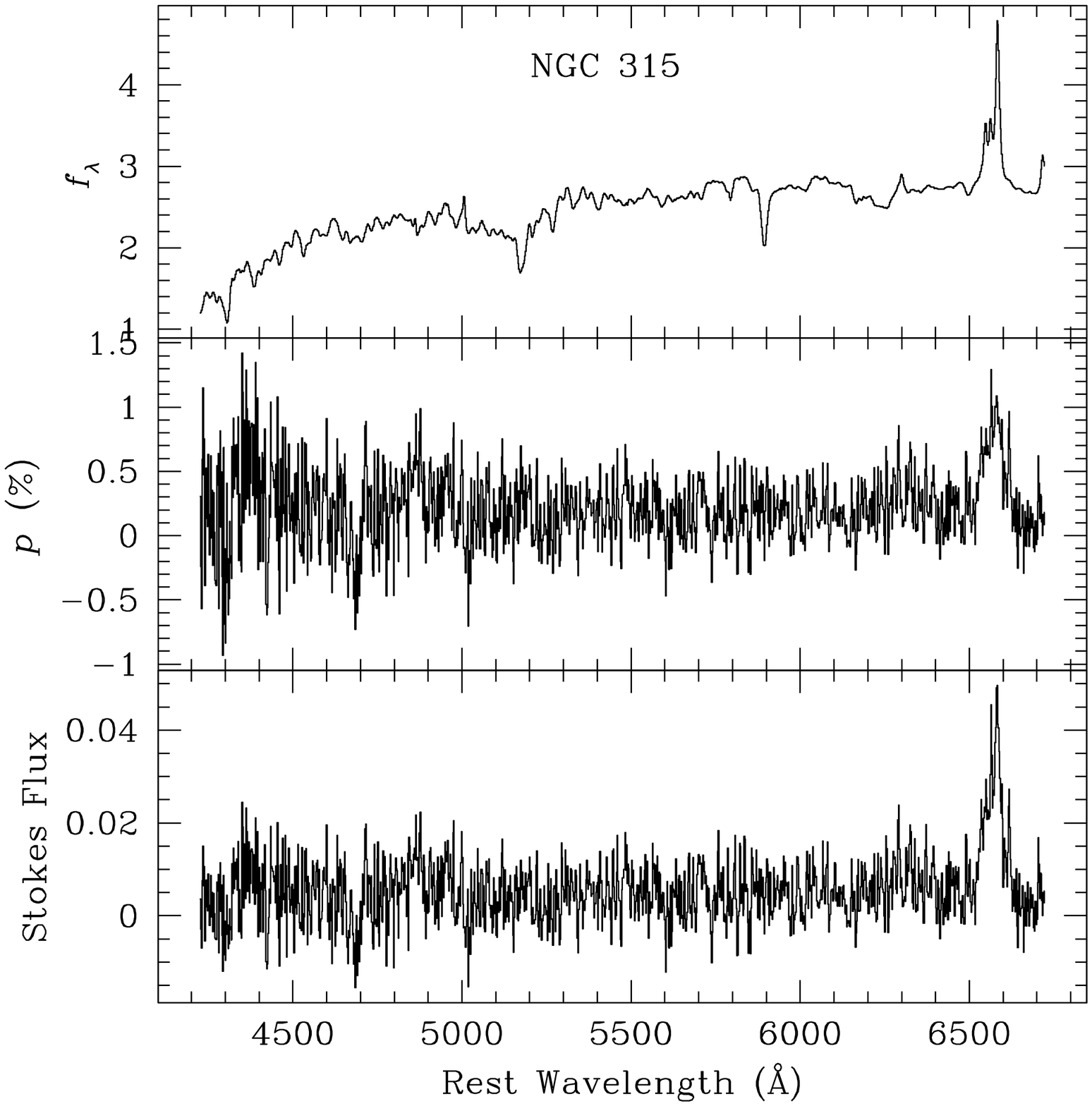}
\caption{Spectropolarimetry data for the LINERs NGC 1052 and NGC 315,
showing polarized broad components of H$\alpha$.  Figures adapted from
Barth et al.\ (1999b,c).
\label{figpol}}
\end{figure}

A follow-up study by Barth et al.\ (1999a) found that the
forbidden-line polarizations in NGC 4258 are correlated with the
critical density of the lines, which can be explained if the disk or
torus surrounds and obscures a substantial portion of a compact,
density-stratified NLR.  Thus, NGC 4258 appears to be a nearly unique
case of a Seyfert 2 with a (partially) hidden NLR seen in scattered
light.  (Partially hidden NLRs have also been detected in some radio
galaxies; see di Serego Alighieri et al.\ 1997.)

The only clear detections of polarized broad-line emission in nearby
LINERs come from a small survey done at the Keck Observatory by Barth,
Filippenko, \& Moran (1999b,c).  This survey targeted 14 LLAGNs,
including LINERs and Seyferts with and without broad H$\alpha$.  Broad
H$\alpha$ polarization was detected in 3 LINERs: NGC 1052, NGC 315,
and (with less certainty) in NGC 4261.  Curiously, these were the only
ellipticals in the survey, and all three are known to have radio jets.
In each case, the angle of polarization of H$\alpha$ is nearly
perpendicular to the jet direction, consistent with the obscuring
torus scenario.  The highest S/N data were obtained for NGC 1052; in
this galaxy the H$\alpha$ emission line has FWHM $\approx5000$ km
s$^{-1}$ in polarized light.  NGC 4261 and NGC 315 are known to have
dusty, 100-pc scale circumnuclear disks, and these are likely to be
the outer extensions of the obscuring structures.  Thus, at least
\emph{some} LINERs have obscuring geometries similar to those of
Seyfert 2s with hidden BLR.  It is worth pointing out, however, that
NGC 1052 and NGC 315 are classified as Type 1.9 objects, since their
broad H$\alpha$ emission was first detected in total-flux spectra.

Polarized emission lines have not yet been detected in any LLAGNs in
spiral galaxies other than NGC 4258.  There are two likely reasons.
Most of the type 1 objects are probably seen directly rather than in
scattered light, so no strong polarization is expected.  Also, the
\emph{HST} surveys have shown that many of the type 2 nuclei lie
behind a thick veil of foreground dust, and this larger-scale
obscuring material would extinguish any polarized light from the
nucleus even if there were an obscuring torus on parsec scales.

\section{What are the Type 2 LINERs and Transition Objects?}

The most important new data constraining the nature of LINER 2 and
transition nuclei have come from recent surveys in the radio and
X-rays; in these spectral regions it is possible to detect central
engines that are completely obscured in the optical and UV.  In a VLA
survey, Nagar et al.\ (2000) find that 64\% of LINER 1s and 36\% of
LINER 2s have compact radio cores; those with multifrequency data are
found to be flat-spectrum sources.  The undetected objects could still
have radio cores which are intrinsically fainter, or undetected due to
free-free absorption.  This result sets a plausible lower limit to the
fraction of genuine AGN central engines in LINER 2s.  Those objects
bright enough for VLBI observations at 5 GHz were studied by Falcke et
al.\ (2000). All showed compact, high-brightness-temperature cores
with $T_b \ga 10^8$ K, confirming that an AGN rather than a starburst
is responsible for the radio emission.  On the other hand, Nagar et
al.\ only detect a compact radio core in 1 out of 18 transition
objects.  Similarly, Filho et al.\ (2000) find a low (20\%) detection
rate of compact radio cores in a sample of 25 transition nuclei.  The
lack of radio cores in most transition objects does not prove that
they are \emph{not} AGNs, however.  As Nagar et al.\ point out, LINERs
show a correlation between radio power and [O~I] luminosity, and
transition objects have systematically lower [O~I] luminosities than
LINERs, so their radio sources are expected to be fainter.

X-ray observations provide another direct probe of the central
engines.  Using \emph{ASCA} spectra, Terashima et al.\ (2000a,b) show
that LINER 1s tend to follow the same correlation traced by
higher-luminosity Seyferts between X-ray and H$\alpha$ flux,
supporting the interpretation of LINER 1s as photoionized AGNs.  LINER
2s have systematically lower X-ray to H$\alpha$ flux ratios, and
Terashima et al.\ conclude that either they contain heavily obscured
AGNs (with $N_H > 10^{23}$ cm$^{-2}$), or they are primarily powered
by stellar processes.  Similar conclusions have been reached by
Roberts et al.\ (2001), using \emph{ASCA} and \emph{ROSAT} data.
Since LINER 2s have very faint X-ray sources, which may be surrounded
by diffuse emission or X-ray binaries of comparable brightness,
observations at high spatial resolution are crucial if nuclear sources
are to be detected.  A \emph{Chandra} survey of 24 LLAGNs by Ho et
al.\ (2001) provides a dramatic new high-resolution view of nearby
galactic nuclei.  In this survey, all of the Type 1 LLAGNs were found
to have nuclear point sources, and 4 out of 5 LINER 2s show compact
nuclear emission as well.  Consistent with the earlier ASCA results,
however, the nuclear X-ray sources in LINER 2s are underluminous in
comparison with LINER 1s for a given H$\alpha$ luminosity. Compact
nuclear X-ray sources were found in only 2 out of 8 transition nuclei.
The overall picture emerging from these observations is that the
majority of LINER 2s are likely to contain AGNs, although their
central engines may be intrinsically fainter than those of LINER 1s.

One slight twist on the possible LINER 1/2 unification is that there
may be LINER 2 nuclei which are AGNs, but which do not contain
obscured Type 1 nuclei.  It is possible to find at least a few nearby
LLAGNs which show no broad H$\alpha$ emission lines in the
highest-quality spectra available, but which also appear to be largely
unobscured based on measurements of the X-ray absorbing column or
observations of the optical/UV featureless continuum. NGC 4594, the
Sombrero galaxy, is a good example (Nicholson et al.\ 1998); another
may be M87, which is a LINER 2 (Ho et al.\ 1997b) that is not hidden
behind an obscuring torus (Whysong \& Antonucci 2001).  These two
galaxies have very massive black holes ($\ga 10^9 M_\odot$) and both
are extremely sub-Eddington accretors, with $\la 10^{-5}
L_{\mathrm{Edd}}$ (Ho 1999).  It is tempting to speculate that at
extremely low \emph{\.{m}}, the BLR fades dramatically or even ceases
to exist altogether; this could be due to a shortage of gas, or the
weakness of the ionizing UV continuum, or a change in the structure of
the accretion flow.  Such ``naked Type 2'' objects would be
preferentially detected in early-type galaxies with very massive black
holes because an AGN with $L < 10^{-5} L_{\mathrm{Edd}}$ would be
extremely faint in a spiral galaxy having $M_{\mathrm{BH}} < 10^8
M_\odot$.

Another possibility is that some LINER 2s may be ``fossil'' AGNs in
which the ionizing continuum of a Type 1 LINER has very recently
turned off.  High-excitation lines such as [O~III] would decay in a
timescale of a few decades while the decay time for lower-excitation
species such as [O~I] and [N~II] would be roughly an order of
magnitude longer (Eracleous, Livio, \& Binette 1995).  Maiolino (2000)
has suggested, along these same lines, that some LINER 2s may be
fossils of bright Seyferts. There must be some objects in the universe
which fit this description, but they probably amount to only a tiny
fraction of the LINER 2 population since the NLR fades so rapidly, and
because LINER 2s are far more numerous than either LINER 1s or bright
Seyferts.  Also, in this scenario the fading [O III]-emitting region
in LINER 2s should appear as a bubble or shell around the nucleus, and
the \emph{HST} imaging survey by Pogge et al.\ (2000) found no
examples of such morphology.

The simplest explanation for the emission-line spectra of transition
nuclei is that they are composite objects in which a LINER is
surrounded by star-forming regions (Ho et al.\ 1993).  In a
ground-based aperture, their spectra would be a mix of AGN and H~II
region emission, accounting for the weakness of [O~I] and other
low-ionization lines.  This scenario also accounts well for the Hubble
type distribution of transition nuclei, which is intermediate between
LINERs (which have a preference for early-type hosts) and H~II nuclei
(most commonly found in Sbc-Sd galaxies).  In such composite systems,
high-resolution spectroscopy can be useful for disentangling the
contributions of the individual components (Gon\c calves et al.\
1999).  However, this interpretation is challenged by the recent
surveys cited above, because compact AGN-like radio and X-ray cores
are found in only a small fraction of transition nuclei.  The lack of
any clear indication of AGN emission in most transition nuclei is a
sufficient reason to consider alternative models for their power
source; it would be dangerous to assume that all transition nuclei
contain AGNs simply because their optical spectra do not resemble
normal H~II regions.  Similar considerations apply to at least part of
the LINER 2 population, because a significant minority of them still
lack conspicuous signs of an AGN central engine.  Furthermore, the few
existing UV spectra of LINER 2s and transition nuclei generally show
that the UV continuum arises from young stars, not an AGN (Maoz et
al.\ 1998).

Along these lines, a variety of models have been proposed to explain
the properties of LINERs and transition objects solely on the basis of
stellar phenomena.  The main challenge for such models is to explain
how the ionized gas surrounding a population of hot stars would emit
enhanced levels of low-ionization forbidden lines, such as [O~I]
$\lambda6300$, in comparison with the spectra of normal H~II regions.
Space limitations preclude a complete discussion of these models so
only a brief listing will be given.  These include models based on
photoionization by hot stars, including O stars with $T_{\mathrm{eff}}
\ga 45,000$ K (Filippenko \& Terlevich 1992; Shields 1992);
photoionization by young starbursts containing Wolf-Rayet stars (Barth
\& Shields 2000); and photoionization by an aging starburst combined
with shock heating from supernova remnants (Engelbracht et al.\ 1998;
Alonso-Herrero et al.\ 2000).  Each of these models can readily
reproduce transition-type optical line ratios, but each is also
subject to potentially serious caveats and it is unlikely that any one
of them can explain the entire transition-object population.  LINER
spectra can be reproduced with some difficulty (i.e., by pushing the
stellar effective temperature or nebular density very high), and it is
probably fair to conclude that these starburst-based models are more
likely to apply to transition objects than ``pure'' LINERs.

Finally, LINER-like emission in ellipticals and spiral bulges can also
be powered by ionizing photons from an evolved stellar population.
Binette et al.\ (1994) demonstrated that post-AGB stars in ellipticals
will produce a dilute, hard ionizing radiation field that should
result in a LINER spectrum in the surrounding gas.  The expected
H$\alpha$ equivalent widths are of order 1 \AA, similar to the levels
observed near the centers of many early-type galaxy bulges.  Planetary
nebula nuclei may also contribute to this effect (Taniguchi et al.\
2000).  Thus, galaxies without either an AGN or very recent star
formation could still be classified as LINERs or transition nuclei
provided that there is sufficient diffuse gas in the nuclear regions
to be detected.

\section{Conclusions}

Traditional unification models based on the orientation of an
optically thick obscuring torus apparently do apply to at least some
fraction of the LLAGN population, as shown by a few detections of
emission-line polarization, ionization cones, and heavily obscured
X-ray sources.  These detections constitute a small minority of the
Type 2 LLAGN population, however.  Obscuration on the scale of the
host galaxy plays a more important role in affecting our view of the
central regions of LINERs, at least in the UV and optical.

The larger question is whether Type 2 LLAGNs contain genuine AGN-like
central engines at all.  Radio and X-ray surveys suggest that the
answer is ``yes'' for many, and perhaps most, LINER 2s.  From the
results of their \emph{Chandra} survey, Ho et al.\ (2001) estimate
that at least 60\% of all LINERs contain AGNs, but this result is
still based on very small-number statistics.  Further observations
with \emph{Chandra} will be the best way to improve on this estimate
and complete the census of nearby AGNs, both by surveying larger
samples and by obtaining deeper exposures to search for obscured or
intrinsically very faint sources.  Transition objects as a class
present the weakest case for being members of the AGN family, and
their spectra may instead be a manifestation of star formation in the
extreme environment of galaxy centers.

\end{document}